# Sodium Trimer Ordering on $Na_xCoO_2$ Surface

Woei Wu Pai[1,*], S. H. Huang[1,2], Ying S. Meng[3], Y. C. Chao[1], C. H. Lin[1], H. L. Liu[2], F. C. Chou[1,4]

[1]Center for Condensed Matter Sciences, National Taiwan University, Taipei 106, Taiwan, R.O.C.

[2]Department of Physics, National Taiwan Normal University, Taipei 106, Taiwan, R.O.C.

[3]Department of Materials Science and Engineering, University of Florida, Gainesville, FL 32611, U.S.A.

[4]National Synchrotron Radiation Research Center, HsinChu 30076, Taiwan, R.O.C.

[*]corresponding author. Email: wpai@ntu.edu.tw

*Abstract*

Sodium ion ordering on *in situ* cleaved $Na_xCoO_2$ ($x$=0.84) surface has been studied by ultra high vacuum scanning tunneling microscopy (UHV-STM) at room temperature. Three main phases, with *p*(3×3), (√7×√7), and (2√3×2√3) hexagonal unit cells and surface Na concentration of 1/3, 3/7, 1/2, respectively, were identified. One surprising finding is that Na trimers act as the basic building blocks that order in long range. The stability of Na trimers is attributed to the increased Na coordination with oxygen as indicated by *ab initio* calculations, and possibly at finite temperature by configuration entropy.

The conducting layered sodium cobaltate, $Na_xCoO_2$, exhibits a series of surprising electronic or magnetic properties with Na concentration $x$[1]. Hydrated $Na_{0.3}CoO_2 \cdot 1.3H_2O$ is superconducting [2]. At $x=0.5$, an unexpected insulating state below ~51 K is observed [3, 4]. For $x>0.5$, the material has a "Curie-Weiss metal" behavior, and A-type antiferromagnetic ordering below ~22 K is found for $x \geqq 0.75$ [1, 5, 6]. $Na_xCoO_2$ also shows large thermopower enhancement at high $x$ [7]. Despite these intense research efforts, exact Na ordering and its subtle role remains largely uncertain, although its importance has been acknowledged [8]. If stable Na ordered patterns exist [4, 9, 10], the electronic or magnetic properties of the $CoO_2$ layers could be affected by, e.g., reconstructed Fermi surface [11, 12], Coulomb potential landscape [13, 14], or coexistence of localized spins and itinerant charges [15]. It is thus important to understand the principle of Na ordering. Here we report the first scanning tunneling microscopy (STM) study of Na ordering on a $Na_xCoO_2$ surface. Our study has one unique advantage, i.e., structure models based on real-space Na arrangement can now be analyzed, shedding new light on the mutual $Na^+$-$Na^+$ and $Na^+$-$Co^{+3/+4}O_2$ interactions. We found a new guiding rule for Na ordering, viz, Na atoms form trimers that act as the basic building blocks for surface Na ordering.

Generally, the driving force of Na ion ordering is thought to be electrostatics [9, 14], which maximizes Na-Na separation. Although many experimental and theoretical studies are thus interpreted, inconsistency is common. Various Na ordered structures observed by electron, X-ray, and neutron diffraction [4, 13, 14, 16] are often at odds with theoretical predictions [9, 10, 13, 14, 17]. This is perhaps due to sample quality variation, Na concentration uncertainty, probe beam induced damage, and computation limitation. The Coulomb repulsion between Na cations can also lead to unintuitive structure. A recent study [14] reveals the formation of ordered vacancies at some simple fractional fillings for $x>0.5$. Inside each vacancy, there exists a Na monomer or a Na trimer occupying Na(1) sites that sit atop Co atoms. We found that this multi-vacancy model and Na trimer geometry explain the observed superstructures of $x=0.84$ and $x=0.71$ crystals very well [15]. In ref. [14], the proposed vacancy models apply to $x\geq 0.5$. For $x<0.5$, Na cations are still presumed to repel each other[9, 14, 17]. In this work, we observed unexpected Na trimer ordering with surface concentration $x_s \leq 0.5$, i.e., $x_s=1/3$, 3/7, and 1/2 with STM. The simple rule of minimizing Na-Na Coulombic repulsion thus fails, indicating an unknown driving force that binds Na ions into clusters. We shall suggest that the Na trimer is stabilized by increased Na coordination with oxygen [8] and possibly at finite temperature by configuration entropy, and the trimer structure is likely relevant to bulk Na ordering.

Bilayer γ-phase $Na_{0.84}CoO_2$ single crystal was grown using an optical floating-zone method and the Na content was confirmed to be (0.84±0.02) through electron probe microanalysis [18]. Cobaltate crystals (~ $2 \times 2 \times 0.5$ mm) were cleaved *in situ* in UHV at room temperature and imaged with an Omicron VTSTM in a chamber described previously [19]. The cleaved surfaces showed steps that are multiples of half unit cell height ($c/2$), with $c$~10.8 Å. This is consistent with the $x$ vs. $c$ relation [1, 20] and the weaker interactions between adjacent $CoO_2$ planes. The ordered Na patterns were consistently observed in ~ 20 samples from two synthesized batches.

Clearly, the surface Na concentration $x_s$ is not necessarily 0.84. In the γ-phase cobaltate, Na atom sits in a trigonal prismatic cage formed with identical oxygen triangles above and below it [21]. Therefore, $x_s$ is expected to be around half of the bulk concentration, i.e., 0.42. Experimentally, we observed three Na ordered patterns, with $p(3\times3)$, ($\sqrt{7}\times\sqrt{7}$), and ($2\sqrt{3}\times2\sqrt{3}$) unit cells (Fig. 1), and proposed concentrations $x_s =1/3$, 3/7, and 1/2 (Fig. 2), respectively. The concentration of 3/7 is very close to half of the bulk concentration and indeed has been most often observed than the other two phases.

Figure 1 presents typical STM micrographs of the three observed phases. Figs. 1(d) – 1(f) are magnified views of Figs. 1(a) – 1(c) phases, respectively. The $p(3\times3)$ phase forms a Kagome lattice with interlaced triangles (Fig. 1(a)). The $(\sqrt{7}\times\sqrt{7})$ phase (Fig. 1(b)) is a simple hexagonal lattice. In Fig. 1(c), the $(2\sqrt{3}\times2\sqrt{3})$ phase is a honeycomb lattice that appears pinwheel-like. The $p(3\times3)$, $(\sqrt{7}\times\sqrt{7})$, and $(2\sqrt{3}\times2\sqrt{3})$ phases will hereafter be denoted as "Kagome", "hex", and "pinwheel" phases. The phases of Fig. 1 are indeed due to surface Na atoms because of (1) the presence and appearance of the phases are uncorrelated to foreign gaseous species contamination; (2) the population of each pattern varies in each cleaved surface; (3) the hex phase can fluctuate with time, showing mobile defects and coexisting mobile "liquid" and "solid" phases (not shown). These results indicate that the observed patterns are related to Na atoms rather than the $CoO_2$ layers or foreign species.

Proposed models for the Kagome, hex, and pinwheel phases are shown in Fig. 2. From Fig. 1(d), we propose the Fig. 2(a) model in which each interlaced green triangle of the $p(3\times3)$ Kagome lattice ($x_s =1/3$) consists of three Na atoms in a trimer unit. All Na atoms in a trimer presumably occupy the most preferred Na(2) site amid a Co triangle. Each trimer can center around either an oxygen or a Na(1) (atop cobalt) atom, hereafter denoted as "O-" or "Co-" centered trimer respectively. We propose Co-centered trimers for the Kagome phase. Note that a Na(1) site is surrounded by the closest Na(2) and oxygen sites arranged in a hexagon and Na atoms do not prefer to adsorb atop oxygen atoms. The Co-centered trimer scenario is thus consistent with the atom-resolved appearance of a "frozen" trimer because Na atoms only occupy the Na(2). Since the observed Kagome lattice has a $p(3\times3)$ unit cell instead of a $p(2\times2)$ or $p(4\times4)$ even periodicity, it cannot be symmetrically commensurate on the $CoO_2$ base lattice. We observed such asymmetry as illustrated by the profile along the dashed line in Fig. 1(d). The Na-Na distance alternates with a ratio of ~ 1.4, smaller than the ratio of 2 if all Na atoms occupy the ideal Na(2) positions. Na atoms thus must have repelled each other toward the two next preferred Na(1) sites, presumably due to Coulomb interaction. Such "Na(2)→Na(1)" displacement, as denoted by the arrows in Fig. 2(a), leads to "expanded" trimers. We also note that in contrast to the suggested ($\sqrt{3}\times\sqrt{3}$) ordering [4, 17], the real space $p(3x3)$ Kagome Na ordering is unexpected and should be re-examined for its potential relevance to the superconductivity in hydrated sodium cobaltate ($x$~1/3), where water layer could serve as a buffer zone to allow surface-like Kagome Na ordering.

The structure model of the hex phase ($x_s =3/7$) consists of ordered trimer clusters, as shown in Fig. 2(b). Our data show that each protruded dot is significantly larger in size than that in the Kagome phase. Each dot can show two or three different topographic contrasts (circle in Fig. 1(b)), suggesting that it is a Na atom cluster rather than a single Na atom. We propose that each dot is an O-centered Na trimer. Due to the ionic radius and Coulomb repulsion of $Na^+$, neighboring Na(1) and Na(2) sites cannot be simultaneously occupied. An O-centered trimer thus occupies either all Na(1) or Na(2) sites. Fig. 1(e) suggests that each trimer rotates concertedly around its center (O site) at room temperature, smearing out resolution of individual Na atoms. We believe that the hex phase trimers are also expanded in size. We also observed dimmer interstitial Na atoms exclusively at the same half of the unit cell (e.g., letter **A** in Fig. 1(e)). This corroborates with the proposed ($\sqrt{7}\times\sqrt{7}$) unit cell and the bright dot as a cluster. Mirror hex domains intersected at ~22° (21.8° predicted) were also observed (Fig. 3(b)). The vertical mirror planes are along {120} and one such plane is denoted as **M** in Fig. 2(b). We note that such mirror domains exist either on the surface or in the bulk; e.g., we have observed ($\sqrt{13}\times\sqrt{13}$) mirror domains for bulk $x$=0.84 using X-ray Laue diffraction [15].

Fig. 2(c) shows the proposed ($2\sqrt{3}\times2\sqrt{3}$) pinwheel phase model ($x_s$ =1/2). In this phase, individual Na atoms were barely resolved, suggesting dynamic switching of occupied Na sites. Each pinwheel segment appears triangle-like (Figs. 1(c) and 1(f)), cf. round dots in Figs. 1(b) and 1(e). To account for such a subtle difference, we assign Co-centered trimers to the pinwheel phase. In contrast to the Kagome phase in which each Na(2) atom can be displaced along two possible Na(2)→Na(1) paths, here the more crowded trimer arrangement favors only one Na(2)→Na(1) path. This leads to the larger triangular shape of each Co-centered trimer, as shown by the dotted triangles in Fig. 2(c).

Na distribution on the cleaved surface is inhomogeneous. Nanoscale hex phase ($x_s$ =3/7) coexists with either the Kagome ($x_s$ =1/3) or the pinwheel ($x_s$ =1/2) phase, as shown in Fig. 3. The Kagome and pinwheel phases have not been found to coexist. Note that $x_s$ =1/3 < 3/7 < 1/2, the phase coexistence is thus natural if $x_s$ =3/7 is the major stable phase between $x_s$ =1/3 and 1/2. This is supported by our electrochemical Na de-intercalation study that shows a stable $x$~3/7 bulk phase does exist within a narrow surface potential range [20]. We caution that the inhomogeneity and nanoscale phase separation can lead to questionable interpretation of photoemission spectroscopy (PES) data [22]. For example, it was suggested that strong surface disorder can eliminate $e_g^{'}$ pockets that have so far evaded probing by PES [23]. Likewise, low energy electron diffraction of Na ordering cannot be readily detected.

One immediate question is whether the Na trimer models are relevant in bulk crystals. STM cannot directly verify this issue. We wish, however, to suggest such relevance by showing that the trimer models can explain two unresolved bulk structure phase transitions in $Na_xCoO_2$. First, neutron powder diffraction reveals a first order structure transition between the so-called H1 and H2 phases for $x \leqq 0.75$ [24]. Na atoms occupy low symmetry *6h* ($2x$, $x$, 1/4) sites at high temperatures (H1 phase) while occupying high symmetry *2c* (2/3,1/3,1/4) sites at low temperature (H2 phase). In our trimer models, whether a trimer consists of all Na(2) or Na(1) atoms, or is O- or Co- centered, the Na atoms at Na(2) sites displace toward Na(1) sites, or vice versa. This places Na at the *6h* sites and always leads to two shorter Na-O bonds and one longer Na-O bond, as observed in neutron studies [24, 25]. In Ref. [24], the *6h* Na occupation is suggested to come from second-nearest-neighbor Na(2)-Na(1) repulsion. Instead, we believe that *6h* sites are caused by nearest-neighbor Na(2)-Na(2) (or Na(1)-Na(1)) repulsion. The *6h* occupation will persist until unfrustrated expanded trimers are fully close-packed. This corresponds to the "TT-stacking" model in ref. [15] and its concentration $x$=0.75 coincides with the H1 and H2 phase boundary. Another unresolved structure transition is found in $x$=0.5 crystals [4, 16]. Transmission electron diffraction (ED) shows hexagonal $p(2\times2)$ superstructure [16] at higher temperature (410 - 470 K) and an orthorhombic ($2\times\sqrt{3}$) supercell at lower temperature [3, 4, 16]. Our pinwheel model can be shown to exhibit bulk $p(2\times2)$ ED pattern if randomized interlayer Na stacking is considered. Furthermore, it can explain the incommensurate transition from the $p(2\times2)$ to ($2\times\sqrt{3}$) phases [4, 16]. Due to space limitation, further details are deferred to a separate publication.

The most salient feature of our Na models is the Na trimer. In fact, we often observed individual trimers on terraces as seen in Fig. 3(a). Why are Na trimers stable if the $Na^+$-$Na^+$ repulsion is significant? One driving force to counterbalance the electrostatic interaction is the reduction of on-site orbital energies of oxygen and cobalt atoms by their coordination with Na atoms. Marianetti et al. [8] have found a simple rule that all low-energy oxygen atoms have three Na nearest neighbors, whereas an oxygen atom with fewer Na neighbors is higher in energy. There are seven oxygen atoms coordinated with an O-centered trimer; one (center oxygen) with three Na and the other six with one Na. There are six oxygen atoms coordinated with a Co-centered trimer; three with two Na and three with one Na. Three separated Na atoms are coordinated with nine oxygen atoms. The tendency of O- and Co- centered Na trimer formation can be attributed to multiple Na coordination with an oxygen atom.

Does the orbital energy reduction suffice to make Na trimers ground state at all temperature? Our preliminary STM data suggest that trimers become unstable upon cooling and reappear at T>250 K. Likewise, the H1 phase [24] (which we argue is due to Na trimer) is observed only at T>320 K($x$~0.75). We therefore investigated the Na trimer stability at T=0 K with *ab initio* calculations using Density Functional Theory within generalized gradient approximations (GGA) with fully relaxed unit cell [10]. We focused on $x$=1/3 since it is a natural fractional concentration with a ($\sqrt{3}\times\sqrt{3}$) unit cell when Na atoms are maximally separated (denoted as "bulk ($\sqrt{3}\times\sqrt{3}$)"). We considered three structures with optimized interlayer Na stacking: (a) bulk ($\sqrt{3}\times\sqrt{3}$) phase; (b) $p(3\times3)$ Kagome phase with O-centered trimers in all layers; (c) $p(3\times3)$ Kagome phase with O- and Co-centered trimers in alternating layers. We found the hybridization of Na and O does lead to energy reduction. Yet, structure (a) is still the ground state at T=0 K. Trimer structures (b) and (c) are respectively ~30 meV and ~40 meV higher in energy per formula unit ($Na_{1/3}CoO_2$). We thus postulate that Na trimers may be stabilized at finite temperature due to configuration entropy, in conjunction with the Na-O orbital energy reduction. As depicted in Figs. 2(a) and (b), an expanded trimer has multiple degrees of freedom to displace a Na(2) atom toward a Na(1) site (entropy gain $\sim k\ln(2)$ per Na), or trimers can rotate around its center concertedly(entropy gain $\sim 1/3\ kT\ln(2)$ per Na). Both lead to free energy reduction at T>0 K. In contrast, the bulk ($\sqrt{3}\times\sqrt{3}$) phase could tolerate little disorder due to the lack of configuration entropy. We note that entropy-driven complex ordering has also been observed in some oxides [26].

This work and ref. [14] indicate the need to verify whether Na trimers do exist in bulk. In particular, the low $x$ regime presents a puzzle because a higher hole concentration does not lead to stronger correlation [8]. We believe that possible Na clustering in bulk at low $x$ may provide useful clues.

In summary, we have discovered surprising Na trimer ordering on $Na_xCoO_2$ surface with STM and proposed realistic structure models. The unexpected trimer stability and its relevance to bulk crystals are discussed. Future STM studies with fine Na concentration and temperature control can hopefully offer further microscopic details of this highly contended and intriguing material. WWP and FCC are supported by NSC-Taiwan. We acknowledge helpful discussion with Prof. P. A. Lee, Mr. Y. Hinuma, Dr. G. J. Shu, Dr. M. Sudip, Dr. M. W. Chu, and Prof. K. C. Lin.

**Figure captions:**

**Fig. 1** (Color online) Ordered Na patterns on cleaved $Na_{0.84}CoO_2$ surface (a-c) and their magnified views (d-f). (a) $p(3\times3)$ "Kagome" phase (16 nm × 16 nm). (b) $(\sqrt{7}\times\sqrt{7})$ "hex" phase (13.5 nm × 13.5 nm). (c) $(2\sqrt{3}\times2\sqrt{3})$ "pinwheel" phase (12.5 nm × 12.5 nm). In each image, the unit cell and a Co-Co close-packed direction are labeled. (d) A line profile shows alternating Na-Na distances. (e) An interstitial Na atom (**A**) is shown. (f) Each pinwheel segment appears triangle-like. Atomic resolution of Na atoms can be discerned. Imaging condition: (a,d) sample bias $V_s$= -1.85 V, tunneling current $I_t$= 70 pA, (b,e) (-2.05 V, 56 pA), (c,f) (-1.78 V, 51 pA).

**Fig. 2** (Color online) Proposed structure models of the (a) Kagome ($x_s$=1/3), (b) hex ($x_s$=3/7), and (c) pinwheel ($x_s$=1/2) phases, which consist of ordered superlattices of Co-, O-, Co-centered Na(2) trimers respectively. In the 2(b) legend, Na(1), Na(2), oxygen sites are denoted. A 3D model is shown in 2(d) (see also, e.g., [24, 25]). All red encapsulated arrows denote the allowed Na(2)↔Na(1) site switching paths. In 2(b), the dotted red circles denote concerted rotation of the Na trimers (green triangles). One substrate mirror plane (**M**), two mirror domains intersected at 21.8°, and an interstitial Na atom (**A**) at a Na(2) site are also shown. In 2(c), each pinwheel segment (trimer) can also "rotate" (see dotted triangles) and renders itself triangle-like (see Fig. 1(f)). For details of each model, see text.

**Fig. 3** (Color online) (a) Coexisting Kagome and hex phases. On the terrace, many individual "trimer" (**T**) dots are also present albeit without clear order (17 × 17 nm$^2$, -2.05 V, 91 pA). (b) Coexisting pinwheel and hex phases. Also note the presence of mirrored hex domains (yellow squares). Both phases are overlaid on a $p(6\times1)$ stripe phase that will be discussed in a separate publication (35 × 35 nm$^2$, -1.78 V, 55 pA).

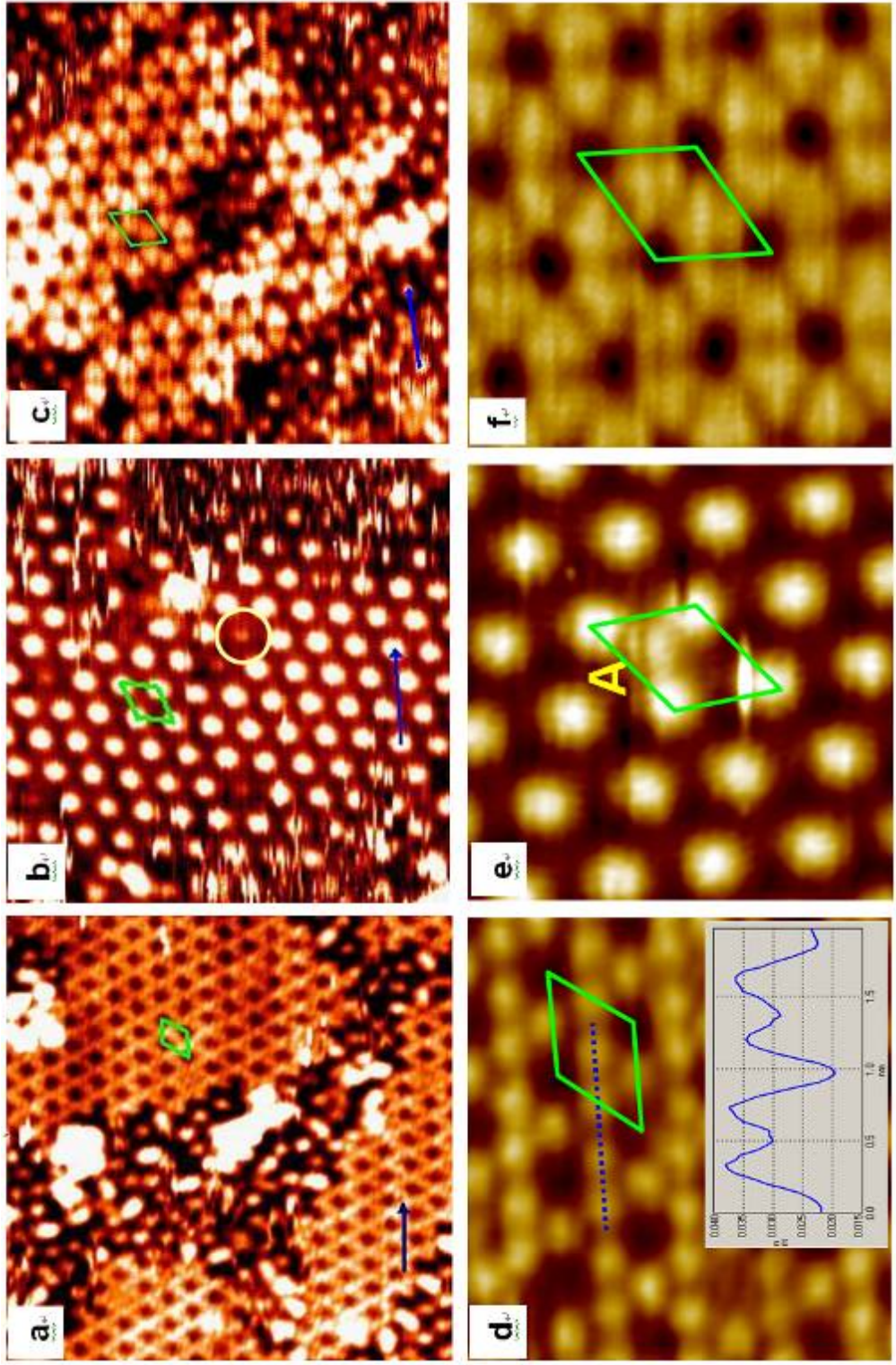

**Figure 1**

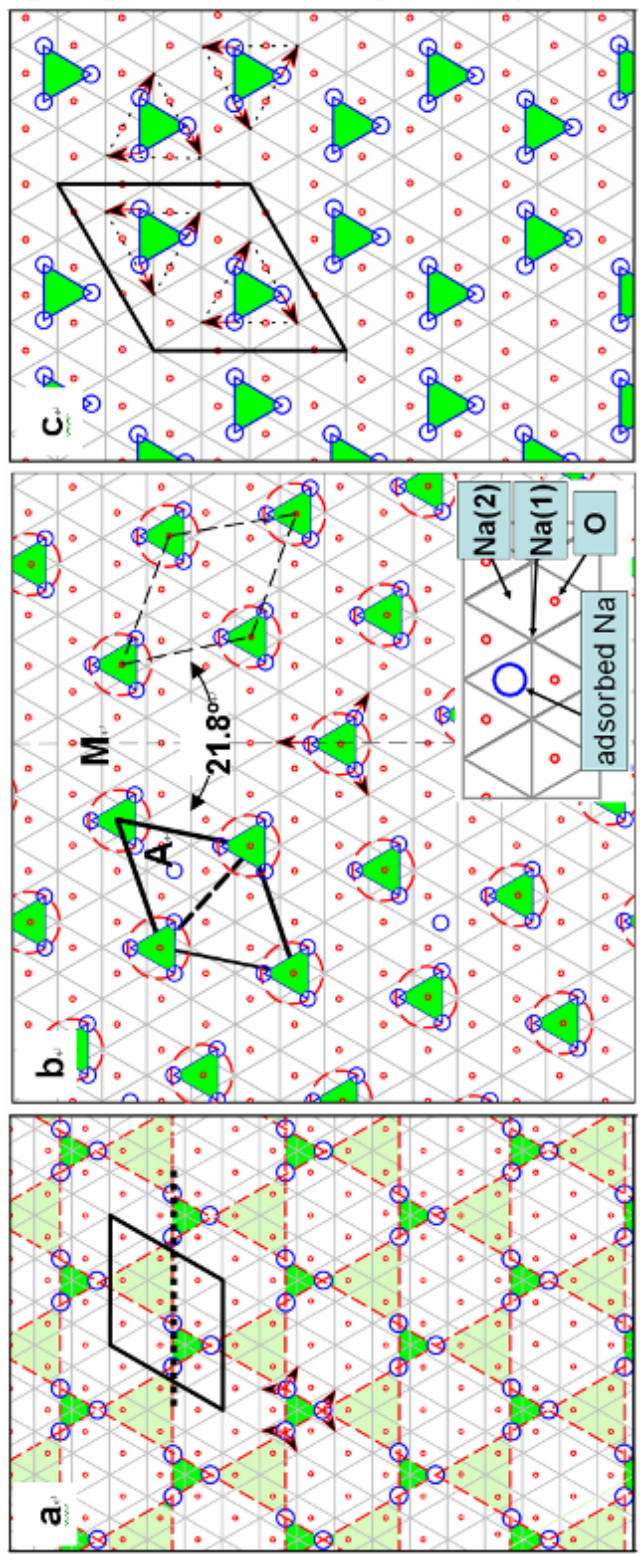

**Figure 2**

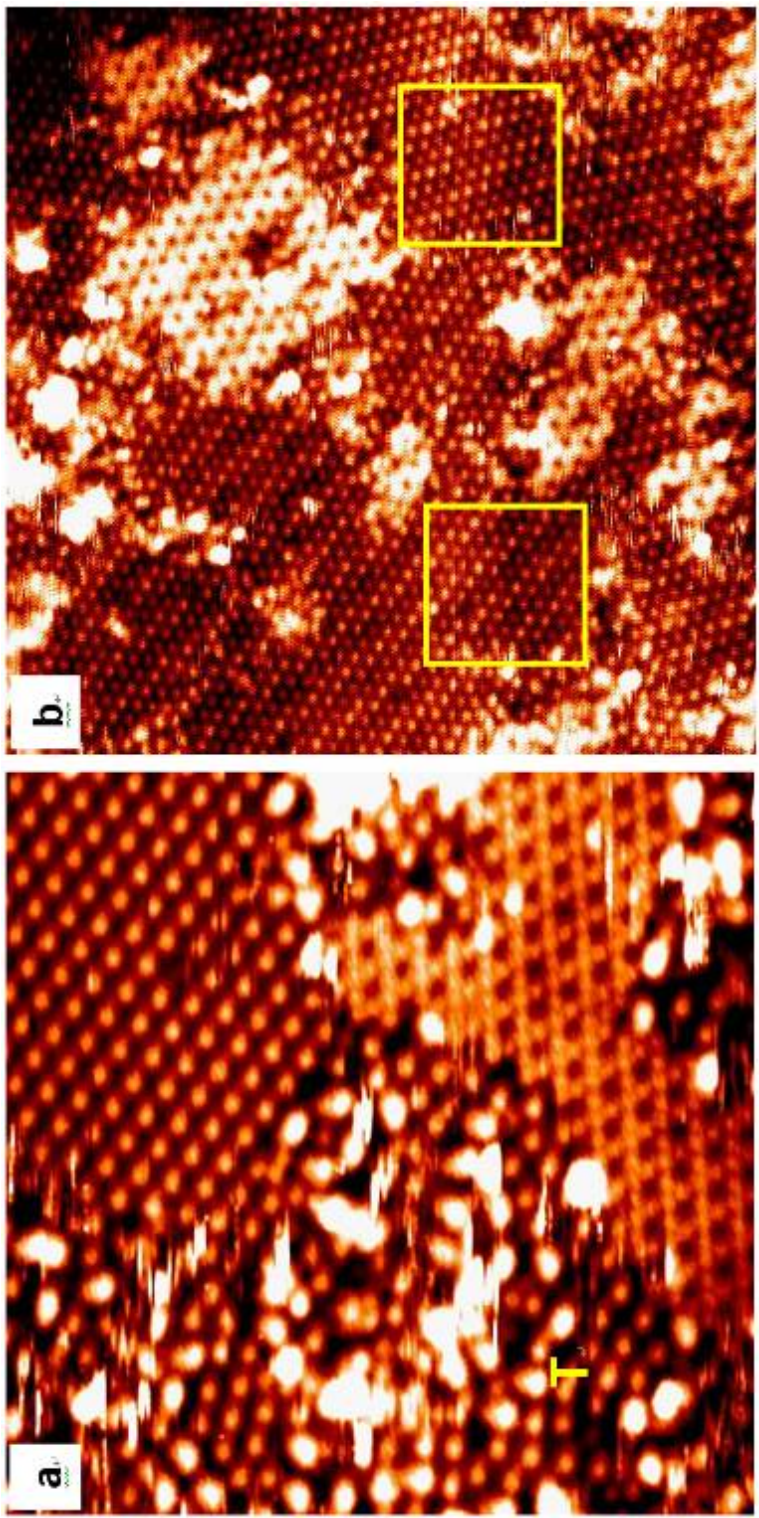

**Figure 3**